\documentclass[aps,prd,showpacs,preprint,nofootinbib]{revtex4-1}
\usepackage{amsmath,amsthm,feynmp,slashed,verbatim,graphicx,color}
\allowdisplaybreaks

\long\def\symbolfootnote[#1]#2{\begingroup%
\def\thefootnote{\fnsymbol{footnote}}\footnote[#1]{#2}\endgroup}

\def\ben{\begin{eqnarray}}
\def\en{\end{eqnarray}}

\begin{document}

\font\el=cmbx10 scaled \magstep2{\obeylines\hfill \today}

\vskip 1.5 cm

\centerline{\large\bf  Anomalous muon magnetic moment in a 
left-right symmetric
}
\centerline{\large\bf composite model}
\small
\vskip 1.0 cm

\centerline{
\bf Piyali Banerjee\symbolfootnote[1]{
banerjee.piyali3@gmail.com
}
}
\medskip
\centerline{\it Department of Physics, Indian Institute of 
Technology Bombay, Mumbai 400076, India}
\bigskip
\bigskip
\begin{center}
{\large \bf Abstract}
\end{center}

We calculate the anomalous magnetic moment for muons
at one loop level arising from left right symmetric excited leptons which
are excited states of known standard model leptons. 
Such excited states arise in compositeness theories where
the known leptons are assumed to be made of more
fundamental particles. In this work, we assume that the excited leptons
possess a left right symmetry also. We show that at the one loop level, 
the QED
contribution to the muon anomalous magnetic moment from these 
excited leptons comes 
from only one Feynman diagram, which turns out to be a natural 
analog of the 
only diagram contributing to anomalous muon magnetic moment in 
the Standard Model.
\medskip
\medskip

PACS Numbers: 
\vfill

\section{Introduction}
In 1948 Schwinger~\cite{Schwinger:1948}, Feynman~\cite{Feynman:1948} and 
Tomonaga~\cite{Tomonaga:1948} showed that a
charged lepton interacting with external electromagnetic field, would 
give rise to a magnetic moment which, in units of Bohr magneton, 
is given by
\[
\mathbf{\mu}_l = g_l 
\left(
\frac{e_l}{2 m_l c}
\right)
\hbar 
\frac{\mathbf{\sigma}}{2}
\]
where $g_{l}$ is the gyromagnetic factor of the lepton. The gyromagnetic factor 
represents the 
relative strength of the intrinsic magnetic dipole moment to the 
strength of the spin-orbit coupling for the lepton. The Dirac equation predicts
$g_l = 2$. The value of 
$g_{l}$ would shift if contributions from loop diagrams
involving QED, weak and strong interactions are taken into account. 
This shift is known as anomalous magnetic moment($a_{l}$) defined as $(g_l-2)/2$.

Several experiments have been conducted to precisely measure the 
anomalous magnetic
moment for electrons and muons. For muons the very precise 
experimental value~\cite{Beringer:2012} of
\[
a_\mu^{\mbox{exp}} = 
11 659 208.9(6.3) \cdot 10^{−10}
\]
is due to the experiment E821 carried out at the Brookhaven laboratories 
\cite{Bennett:2006,Roberts:2010}.
The experimental results are at about $3\sigma$ away from the Standard 
Model (SM) theoretical predictions \cite{Jegerlehner:2009}.
New experiments are underway at Fermilab \cite{Roberts:2010, Carey:2009} 
and at J-PARC \cite{Mibe:2010} to confirm
the above value and reduce the experimental uncertainty. In order to 
reach an accuracy comparable to experimental results theoretical 
calculations 
of $a_{\mu}$ have been reviewed and revisited \cite{Knecht:2014}.  
Thus considering both the experimental result and the SM calculation on
equally firm footings, 
the $3\sigma$ discrepancy clearly hints at new physics beyond SM. 
Several new models aiming to
address various shortcomings of SM may be able to
explain the discrepancy, at least in part. For example, many Beyond Standard
Model theories
which predict a new spectrum of fermions also contribute
to the magnetic moment for known leptons at one loop level. At present
in the absence of any direct evidence of beyond
Standard Model particles, the $g-2$ experimental results
serve to constrain the parameter space of such models.

Composite models predict new spectra of fermions which
can contribute to the anomalous magnetic moment of the known leptons.
In these models known SM 
fermions are not treated as point like but composite \cite{Baur:1990},
thereby predicting a rich spectrum of excited fermions.
The known
fermions can be thought of as the ground state of these excited fermions.
Excited fermions can contribute a large anomalous magnetic moment to
ordinary SM leptons, as investigated in several works 
\cite{Terazawa:1982,Renard:1982, Rakshit:2001}.
However, these works assume that excited leptons only obey left handed
symmetry just like ordinary leptons. 

The confirmation of small neutrino masses 
\cite{Fukuda:2001nk,Ahmad:2002jz,Ahmad:2002ka,Bahcall:2004mz} in 
1998 has strengthened
the possibility that the spectrum of matter is after all symmetric between 
the two handedness states, with the maximal parity violating weak forces 
being a low energy effect. The Left-Right symmetric model 
\cite{Pati:1974yy,Mohapatra:1974gc} through
the see-saw mechanism \cite{Minkowski:1977sc,GellMann:1980vs,Yanagida:1979as} 
can explain the small neutrino masses \cite{Mohapatra:1979ia}. 
In this model both left and right handed fermions are taken as doublets.
If the excited lepton
spectrum together with its ground state has both chiralities, that
is, if there are both left and right handed leptons along with their
excited states \cite{Banerjee:2014}, then the contribution to the 
anomalous magnetic moment of ordinary SM leptons will be different. 
The contribution to $a_{\mu}$ from the spectrum of the left right symmetric
composite model can be
used to explain the experimental measurements and hence can be used to
constrain its parameter space.

\section{Left-right symmetric excited lepton model}
\label{sec:setup}

For the left-right symmetric composite model of \cite{Banerjee:2014}
the magnetic transition between
ordinary lepton and the excited lepton is given by
\begin{equation}
\label{eq:LRLag}
\mathcal{L}_{\mbox{trans}} =
\frac{1}{2\Lambda} \bar{l}^*_L \sigma^{\mu \nu}
\left[
g_s f_s \frac{\lambda^a}{2} G^a_{\mu \nu} +
g_{1} f_{1} \frac{\tau}{2} \cdot \mathbf{W}^{L}_{\mu \nu} +
g_{2} f_{2} \frac{\tau}{2} \cdot \mathbf{W}^{R}_{\mu \nu} +
g'' f'' \frac{B-L}{2} B^{B-L}_{\mu \nu}
\right]
l_R +
\mbox{H.c.}
\end{equation}
Here $W^{L}_{\mu\nu}$ and $W^{R}_{\mu\nu}$ are the field strength 
tensors of $SU(2)_{L}$ and $SU(2)_{R}$ gauge fields
respectively and $B^{B-L}_{\mu\nu}$
is the field strength tensor of $U(1)_{B-L}$. 
$g_{1}$, $g_{2}$ and $g''$
are the $SU(2)_L$, the $SU(2)_R$ and the $U(1)_{B-L}$ gauge couplings respectively. 
Consistent with the original left-right symmetry philosophy we assume $g_1 = g_2$.
$f_{1}$, $f_{2}$ and $f^{''}$ are the new couplings that arise due to
compositeness in the theory. 
The possible gauge mediated transitions
between ordinary leptons and excited fermions of
both chiralities arising from the above Lagrangian are shown in 
Fig.~\ref{fig:newtransitions}. 
\newcommand{\gauge}{g,\gamma,W,Z,W',Z'}
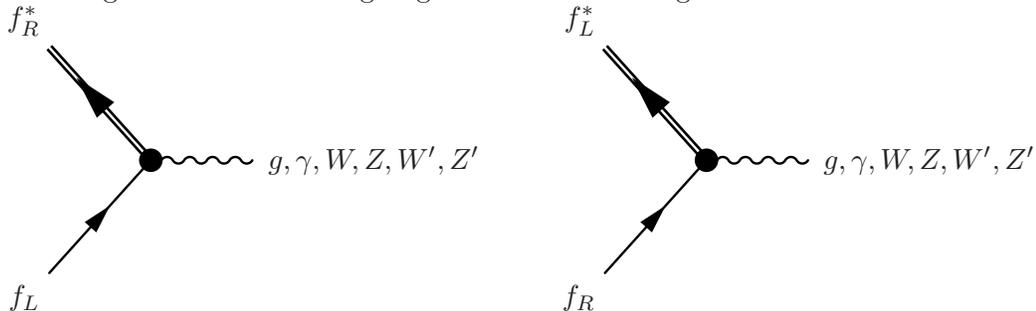
\begin{figure}[!hhh]
\ 

\medskip

\hspace*{-1cm}
\begin{minipage}[c]{7cm}
\begin{fmffile}{newtransitions1}
\unitlength = 1mm
\begin{fmfgraph*}(30,30)
\fmfleft{i1,i2}
\fmfright{o1}
\fmfv{label=$f^*_R$}{i2}
\fmfv{label=$f_L$}{i1}
\fmfv{label=$\gauge$}{o1}
\fmfv{decor.shape=circle,decor.filled=full,decor.size=4thick}{v1}
\fmf{heavy}{v1,i2}
\fmf{fermion}{i1,v1}
\fmf{photon,tension=2}{v1,o1}
\end{fmfgraph*}
\end{fmffile}
\end{minipage}
~
\begin{minipage}[c]{7cm}
\begin{fmffile}{newtransitions2}
\unitlength = 1mm
\begin{fmfgraph*}(30,30)
\fmfleft{i1,i2}
\fmfright{o1}
\fmfv{label=$f^*_L$}{i2}
\fmfv{label=$f_R$}{i1}
\fmfv{label=$\gauge$}{o1}
\fmfv{decor.shape=circle,decor.filled=full,decor.size=4thick}{v1}
\fmf{heavy}{v1,i2}
\fmf{fermion}{i1,v1}
\fmf{photon,tension=2}{v1,o1}
\end{fmfgraph*}
\end{fmffile}
\end{minipage}

\medskip

\caption{
\label{fig:newtransitions}
Transitions between ordinary and left-right symetric fermions via
gauge boson emission.
}
\end{figure}

When spinors couple to photons it gives rise to magnetic 
dipole moment. 
Part of the discrepancy between experimental and SM theoretical 
predictions of muon anomalous magnetic moment may be explained 
by the contribution of left right symmetric excited muons via
photon emission. 
Even though in principle excited states with 
higher spin and various isospin values might exist and contribute
to the anomalous muon magnetic moment, there is to date
no compelling reason propsed for any fine 
cancellation between the contributions from the first excited state 
and other excited states. 
Hence, for a conservative order of magnitude 
estimation we will
consider only the contribution of the lowest lying excited state of 
spin and isospin $1/2$ at the one loop level, and obtain 
constraints on the couplings $f_1$, $f_2$, $f''$ by comparing it with
the discrepancy between theoretical SM and experimental results.

\section{Magnetic moment for left right symmetric model }
\label{sec:setup}
The contribution of excited muons to the anomalous magnetic moment of 
ordinary muons at one loop level 
occurs from the Feynman graphs shown in 
Table~\ref{table:matrixelement}. 
Table~\ref{table:matrixelement} also
lists the corresponding Feynman integrals. In the simplest proposals, 
the effective dipolar coupling for excited leptons to leading order 
in derivatives is introduced through a
dipolar form factor for the excited leptons, given by 
$\frac{\Lambda^4}{(q^2 - \Lambda^2)^2}$ \cite{Renard:1982,Rakshit:2001}, 
where $q^2$ is the virtual photon mass squared. 
We tabulate the contributions to the form factor, $F_{2}(0)$, 
for all the leading graphs
in Tables~\ref{table:anomalous1}, \ref{table:anomalous2}, 
\ref{table:anomalous3}, \ref{table:anomalous4}. 

However, except for graph~1 of Table~\ref{table:matrixelement} 
all other graphs  have their corresponding
mirror images. The contributions to matrix element from the
Feynman graphs 2, 3 and 4 is exactly opposite to those from their
mirror images 2', 3' and 4'. This is nothing but a manifestation of 
the gauge invariance of the underlying
theory. It is interesting to note that the only non-zero SM 
contribution to anomalous 
muon magnetic moment at the one loop level
comes from the SM analog of graph~1. In our left-right symmetric theory 
barring the form factor for
excited lepton and the effective charge nothing changes in the expression 
for matrix element. 
Since in our left-right symmetric theory the only contribution to muon
anomalous magnetic moment that survives comes from graph~1,
we now proceed to provide some details of the calculation of its
contribution to $F_2(0)$.

Employing the Feynman rules on graph~1 we get,
\begin{eqnarray*}
i M^\mu 
& = &
\int \frac{d^4 k}{(2\pi)^4} \, 
\frac{-i g^{\nu \alpha}}{k^2+i\epsilon} 
\bar{u}(q_2) 
(e_{{\rm eff}} k_\beta \sigma^{\nu\beta})
\frac{1}{(1-\frac{k^2}{\Lambda^2})^2}
\frac{
i(\slashed{q_2} + \slashed{k} + M)
}
{
(q_2+k)^2 - M^2 + i\epsilon
}
(-i e \gamma^\mu) \\
&   &
~~~~~~~~~~~~~~~~~~
\frac{
i (\slashed{q_1} + \slashed{k} + M)
}
{
(q_1+k)^2 - M^2 + i\epsilon
}
(e_{{\rm eff}} k_\beta\sigma^{\alpha\beta})
\frac{1}{(1-\frac{k^2}{\Lambda^2})^2} 
u(q_1),
\end{eqnarray*}
where
$\Lambda$ is  the compositeness
scale, $m$ the mass of the ordinary lepton and $M$ the mass of the 
excited lepton.
The term $e_{{\rm eff}}$ above is the effective charge which turns out to be
\[
e_{{\rm eff}} = \frac{e}{\Lambda}(f_1 + f_2 + f''),
\]
where $f_1$, $f_2$ and $f''$ are the couplings present in the left-right
symmetric theory of compositeness.

Using standard techniques for evaluating Feynman integrals, it can
be shown that in the regime $m \ll M \leq \Lambda$, the contribution
to the magnetic moment obeys the following equation.
\begin{eqnarray*}
F_2(0)
& = &
\frac{\alpha}{2\pi} \cdot \frac{8Mm }{\Lambda^2} (f_1 + f_2 + f'')^2
\int_0^1 dt \\
&         & 
\left(
\frac{\frac{t}{2} - \frac{t^2}{3}}{1 - t + t M^2 \Lambda^{-2}} +
\frac{3t^2(1-t) M^2 \Lambda^{-2}}{1 - t + t M^2 \Lambda^{-2}} -
\frac{3 t^3 (1-t) M^4\Lambda^{-4}}{2 (1 - t + t M^2 \Lambda^{-2})^2} +
\frac{t^4(1-t) M^6\Lambda^{-6}}{3 (1 - t + t M^2 \Lambda^{-2})^3} -
\frac{11 t(1-t)}{6}
\right) \\
&   &
{} +
\mbox{lower order}.
\end{eqnarray*}
Above, the phrase ``lower order'' means that the values of the remaining
summands are at least a factor of $1/M$ lower than the given expression.

The above integral can be evaluated explicitly using standard methods to get
\begin{eqnarray*}
F_2(0) 
& = &
(f_1 + f_2 + f'')^2 \cdot \frac{\alpha}{2\pi} \cdot 8 Mm \Lambda^{-2} 
\left(
\frac{1}{2} I_1 + 
\left(3M^2\Lambda^{-2} - \frac{1}{3}\right) I_2 -
3 M^2\Lambda^{-2} I_3 -
\frac{3 M^4\Lambda^{-4}}{2} I_4  
\right. \\
&         &
~~~~~~~~~~~~~~~~~~~~~~~~~~~~~~~~~~~~~~~~~
\left.
{} +
\frac{3 M^4\Lambda^{-4}}{2} I_5 +
\frac{M^6\Lambda^{-6}}{3} I_6 -
\frac{M^6\Lambda^{-6}}{3} I_7 -
\frac{11}{36}
\right),
\end{eqnarray*}
where
\begin{eqnarray*}
I_1 
& = &
\frac{1}{(1-M^2\Lambda^{-2})^2}
(-\ln(M^2\Lambda^{-2}) - (1-M^2\Lambda^{-2})),
\end{eqnarray*}
\begin{eqnarray*}
I_2
& = &
\frac{1}{(1-M^2\Lambda^{-2})^3}
\left(
-\ln(M^2\Lambda^{-2}) 
- 2(1-M^2\Lambda^{-2}) 
+ \frac{1 - M^4\Lambda^{-4}}{2}
\right),
\end{eqnarray*}
\begin{eqnarray*}
I_3
& = &
\frac{1}{(1-M^2\Lambda^{-2})^4}
\left(
-\ln(M^2\Lambda^{-2}) 
- 3(1-M^2\Lambda^{-2}) 
+ \frac{3(1 - M^4\Lambda^{-4})}{2}
- \frac{1-M^6\Lambda^{-6}}{3}
\right),
\end{eqnarray*}
\begin{eqnarray*}
I_4
& = &
\frac{1}{(1-M^2\Lambda^{-2})^4}
\left(
(M^{-2}\Lambda^2 - 1)
+ 3\ln(M^2\Lambda^{-2}) 
+ 3(1-M^2\Lambda^{-2}) 
- \frac{(1 - M^4\Lambda^{-4})}{2}
\right),
\end{eqnarray*}
\begin{eqnarray*}
I_5
& = &
\frac{1}{(1-M^2\Lambda^{-2})^5}
\left(
(M^{-2}\Lambda^2 - 1)
+ 4\ln(M^2\Lambda^{-2}) 
+ 6(1-M^2\Lambda^{-2}) 
- 2(1 - M^4\Lambda^{-4})
+ \frac{1-M^6\Lambda^{-6}}{3}
\right),
\end{eqnarray*}
\begin{eqnarray*}
I_6
& = &
\frac{1}{(1-M^2\Lambda^{-2})^5}
\left(
\frac{M^{-4}\Lambda^4 - 1}{2}
- 4 (M^{-2}\Lambda^2 - 1)
- 6\ln(M^2\Lambda^{-2}) 
- 4(1-M^2\Lambda^{-2}) 
+ \frac{1 - M^4\Lambda^{-4}}{2}
\right),
\end{eqnarray*}
\begin{eqnarray*}
I_7
& = &
\frac{1}{(1-M^2\Lambda^{-2})^6}
\left(
\frac{M^{-4}\Lambda^4 - 1}{2}
- 5 (M^{-2}\Lambda^2 - 1)
- 10\ln(M^2\Lambda^{-2}) 
- 10(1-M^2\Lambda^{-2}) 
\right. \\
&   &
\left.
{} +
\frac{5(1 - M^4\Lambda^{-4})}{2}
- \frac{1 - M^6\Lambda^{-6}}{3}
\right).
\end{eqnarray*}

Note that when $M \uparrow \Lambda$, 
$I_1 \rightarrow 1/2$, 
$I_2 \rightarrow 1/3$, 
$I_3 \rightarrow 1/4$, 
$I_4 \rightarrow 1/4$, 
$I_5 \rightarrow 1/5$, 
$I_6 \rightarrow 1/5$, 
$I_7 \rightarrow 1/6$,
implying that $I_1, \ldots, I_7$ are continuous functions of $M$
when $M = \Lambda$.
We thus get
\[
\mbox{When $M = \Lambda$}: ~~~~~
F_2(0) =
(f_1 + f_2 + f'')^2 \cdot \frac{\alpha}{2\pi} \cdot \frac{m}{\Lambda} \cdot 
\frac{7}{45} +
\mbox{lower order}.
\]

We show some of the steps of the calculation of the Feynman integral
 in detail, for the contributing
graph, in Appendix~\ref{app:calculation}.

\section{Constraint on $|f_1+f_2+f''|$}
\begin{figure}[!hhh]
\includegraphics[width=10cm]{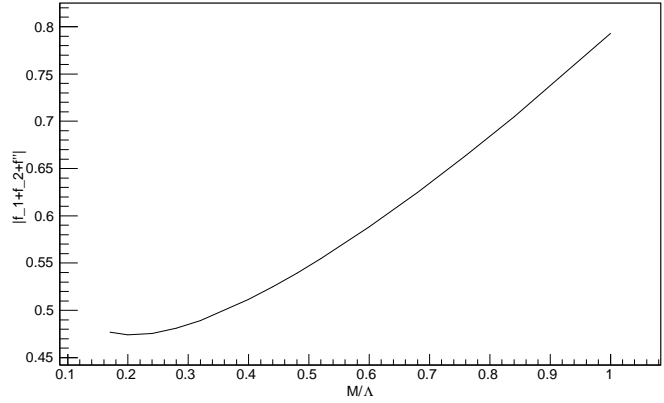}
\caption{
\label{fig:constraint}
At $\Lambda = 5000 ~ \mbox{GeV}/c^2$, for a given value of the 
ratio $M/\Lambda$, the absolute value of the sum of the couplings
 $|f_1+f_2+f''|$ 
has to be below the graph shown.
}
\end{figure}

Since the experimental value of the anomalous muon magnetic moment
exceeds the SM prediction, we can use the experiment minus theory gap
as an upper bound on the QED one-loop contribution due to presence of
left right symmetric excited muons. Thus, we get that
\begin{eqnarray*}
a_\mu^{\mbox{exp}} −- a_\mu^{\mbox{SM}}
& \geq &
\frac{\alpha}{2\pi}
(f_1 + f_2 + f'')^2 \cdot 8 Mm \Lambda^{-2} 
\left(
\frac{1}{2} I_1 + 
\left(3M^2\Lambda^{-2} - \frac{1}{3}\right) I_2 -
3 M^2\Lambda^{-2} I_3 -
\frac{3 M^4\Lambda^{-4}}{2} I_4  
\right. \\
&         &
~~~~~~~~~~~~~~~~~~~~~~~~~~~~~~~~~~~~~~~~~
\left.
{} +
\frac{3 M^4\Lambda^{-4}}{2} I_5 +
\frac{M^6\Lambda^{-6}}{3} I_6 -
\frac{M^6\Lambda^{-6}}{3} I_7 -
\frac{11}{36}
\right),
\end{eqnarray*}
where the terms $I_1, \ldots, I_7$ are functions of $M/\Lambda$ and 
have been defined earlier. Recall that the assumption behind the 
above expression on the right hand side was that $m \ll M \leq \Lambda$.

There are two current `best' determinations of the experiment minus 
theory gap for anomalous muon magnetic moment \cite{Knecht:2014} viz. 
\[
a_\mu^{\mbox{exp}} −- a_\mu^{\mbox{SM}} = 2.37 \times 10^{−9} 
~~~~~
\mbox{OR}
~~~~~
a_\mu^{\mbox{exp}} −- a_\mu^{\mbox{SM}} = 2.74 \times 10^{−9}.
\]
In order to be conservative, we will use the first value for the
gap in the calculation above.
We will also the most precise value of $\alpha$ viz.
$\alpha^{-1} = 137.035999037$~\cite{Bouchendira:2011}, and the most 
precise value of muon mass viz. $m = 0.1056583715 ~ \mbox{GeV}/c^2$ 
\cite{Beringer:2012} known currently. Doing so gives us,
\begin{eqnarray*}
6.11518 \times 10^{-8}
& \geq &
(f_1 + f_2 + f'')^2 \cdot  M \Lambda^{-2} 
\left(
\frac{1}{2} I_1 + 
\left(3M^2\Lambda^{-2} - \frac{1}{3}\right) I_2 -
3 M^2\Lambda^{-2} I_3 -
\frac{3 M^4\Lambda^{-4}}{2} I_4  
\right. \\
&         &
~~~~~~~~~~~~~~~~~~~~~~~~~~~~~~~~~~~~~~~~~
\left.
{} +
\frac{3 M^4\Lambda^{-4}}{2} I_5 +
\frac{M^6\Lambda^{-6}}{3} I_6 -
\frac{M^6\Lambda^{-6}}{3} I_7 -
\frac{11}{36}
\right),
\end{eqnarray*}
For $\Lambda = 5000 ~ \mbox{GeV}/c^2$, the resulting constraint on
the absolute value of the sum of the couplings
$|f_1+f_2+f''|$ has been plotted in
Fig.~\ref{fig:constraint} as a function of $M/\Lambda$ and found 
to be modest, of order 1.

\section{Conclusion}
In this note, we have calculated in detail the QED contribution to 
anomalous 
magnetic moment for muons
at one loop level arising from left right symmetric excited muons which
are the first excited states of known standard model muons. 
As a result of the gauge invariance of the underlying theory,
we have shown that at the one loop level 
only one Feynman diagram contributes, simplifying our calculations. 
We obtain constraints on the parameter space of the left right symmetric
composite model of \cite{Banerjee:2014}, which are similar in spirit
to the constrains obtained earlier by \cite{Renard:1982} for 
left handed excited muons.

\noindent
\begin{table}
\begin{minipage}[t]{7cm}
\begin{center}
Diagram
\end{center}
\end{minipage}
\begin{minipage}[b]{8.5cm}
\begin{center}
$iM^\mu$
\end{center}
\end{minipage}

\medskip

\noindent
1.
\begin{minipage}[t]{7cm}
\begin{fmffile}{muonmagmoment1}
\unitlength = 1mm
\begin{fmfgraph*}(70,20)
\fmfleft{i1,i2}
\fmfright{o1,o2}
\fmfv{decor.shape=circle,decor.filled=full,decor.size=2thick}{v1}
\fmfv{decor.shape=circle,decor.filled=full,decor.size=2thick}{v3}
\fmfv{}{v2}
\fmf{fermion,label=$\ell(q_1)$}{i1,v1}
\fmf{fermion,label=$\ell^*$,label.side=left}{v1,v2}
\fmf{fermion,label=$\ell^*$,label.side=left}{v2,v3}
\fmf{fermion,label=$\ell(q_2)$}{v3,o1}
\fmf{phantom}{i2,v4,o2}
\fmf{photon,tension=0,label=$p$}{v4,v2}
\fmf{photon,right=0.5,tension=0,label=$k$}{v1,v3}
\end{fmfgraph*}
\end{fmffile}
\end{minipage}
\begin{minipage}[c]{8.5cm}
\[
\begin{array}{c}
-\frac{e e_{{\rm eff}}^2}{4} \int 
\frac{d^4 k}{(2\pi)^4} \,
\frac{
\bar{u}(q_2) 
(\gamma^\nu \slashed{k} - \slashed{k} \gamma^\nu)
(\slashed{q_2} + \slashed{k} + M)
\gamma^\mu
(\slashed{q_1} + \slashed{k} + M)
(\gamma_\nu \slashed{k} - \slashed{k} \gamma_\nu)
u(q_1)
}
{
(k^2+i\epsilon)
((k+q_1)^2-M^2+i\epsilon)
((k+q_2)^2-M^2+i\epsilon)
(k^2\Lambda^{-2} - 1)^4
}
\end{array}
\]
\end{minipage}

\medskip

\noindent
2.
\begin{minipage}[t]{7cm}
\begin{fmffile}{muonmagmoment2}
\unitlength = 1mm
\begin{fmfgraph*}(70,20)
\fmfleft{i1,i2}
\fmfright{o1,o2}
\fmfv{decor.shape=circle,decor.filled=full,decor.size=2thick}{v1}
\fmfv{decor.shape=circle,decor.filled=full,decor.size=2thick}{v2}
\fmfv{}{v2}
\fmf{fermion,label=$\ell(q_1)$}{i1,v1}
\fmf{fermion,label=$\ell^*$,label.side=left}{v1,v2}
\fmf{fermion,label=$\ell$,label.side=left}{v2,v3}
\fmf{fermion,label=$\ell(q_2)$}{v3,o1}
\fmf{phantom}{i2,v4,o2}
\fmf{photon,tension=0,label=$p$}{v4,v2}
\fmf{photon,right=0.5,tension=0,label=$k$}{v1,v3}
\end{fmfgraph*}
\end{fmffile}
\end{minipage}
\begin{minipage}[c]{8.5cm}
\[
\begin{array}{c}
-\frac{e e_{{\rm eff}}^2}{4(p^2\Lambda^{-2}-1)^2} \int 
\frac{d^4 k}{(2\pi)^4}  \\
\frac{
\bar{u}(q_2) 
\gamma^\nu
(\slashed{q_2} + \slashed{k} + M)
(\gamma^\mu \slashed{p} - \slashed{p} \gamma^\mu)
(\slashed{q_1} + \slashed{k} + M)
(\gamma_\nu \slashed{k} - \slashed{k} \gamma_\nu)
u(q_1)
}
{
(k^2+i\epsilon)
((k+q_1)^2-M^2+i\epsilon)
((k+q_2)^2-M^2+i\epsilon)
(k^2\Lambda^{-2} - 1)^2
}
\end{array}
\]
\end{minipage}

\medskip

\noindent
2'.
\begin{minipage}[t]{7cm}
\begin{fmffile}{muonmagmoment2prime}
\unitlength = 1mm
\begin{fmfgraph*}(70,20)
\fmfleft{i1,i2}
\fmfright{o1,o2}
\fmfv{decor.shape=circle,decor.filled=full,decor.size=2thick}{v2}
\fmfv{decor.shape=circle,decor.filled=full,decor.size=2thick}{v3}
\fmfv{}{v2}
\fmf{fermion,label=$\ell(q_1)$}{i1,v1}
\fmf{fermion,label=$\ell^*$,label.side=left}{v1,v2}
\fmf{fermion,label=$\ell$,label.side=left}{v2,v3}
\fmf{fermion,label=$\ell(q_2)$}{v3,o1}
\fmf{phantom}{i2,v4,o2}
\fmf{photon,tension=0,label=$p$}{v4,v2}
\fmf{photon,right=0.5,tension=0,label=$k$}{v1,v3}
\end{fmfgraph*}
\end{fmffile}
\end{minipage}
\begin{minipage}[c]{8.5cm}
\[
\begin{array}{c}
-\frac{e e_{{\rm eff}}^2}{4(p^2\Lambda^{-2}-1)^2} \int 
\frac{d^4 k}{(2\pi)^4}  \\
\frac{
\bar{u}(q_2) 
(\gamma^\nu \slashed{k} - \slashed{k} \gamma^\nu)
(\slashed{q_2} + \slashed{k} + M)
(\gamma^\mu \slashed{p} - \slashed{p} \gamma^\mu)
(\slashed{q_1} + \slashed{k} + M)
\gamma_\nu
u(q_1)
}
{
(k^2+i\epsilon)
((k+q_1)^2-M^2+i\epsilon)
((k+q_2)^2-M^2+i\epsilon)
(k^2\Lambda^{-2} - 1)^2
}
\end{array}
\]
\end{minipage}

\medskip

\noindent
3.
\begin{minipage}[t]{7cm}
\begin{fmffile}{muonmagmoment3}
\unitlength = 1mm
\begin{fmfgraph*}(70,20)
\fmfleft{i1,i2}
\fmfright{o1,o2}
\fmfv{decor.shape=circle,decor.filled=full,decor.size=2thick}{v1}
\fmfv{}{v2}
\fmfv{decor.shape=circle,decor.filled=full,decor.size=2thick}{v3}
\fmf{fermion,label=$\ell(q_1)$}{i1,v1}
\fmf{phantom}{i2,w1}
\fmf{photon,tension=0,label=$p$}{w1,v1}
\fmf{fermion,label=$\ell^*$,label.side=right}{v1,v2}
\fmf{phantom}{w1,w2}
\fmf{photon,left=0.8,tension=0,label=$k$}{v2,v3}
\fmf{fermion,label=$\ell^*$,label.side=right}{v2,v3}
\fmf{phantom}{w2,w3}
\fmf{fermion,label=$\ell(q_2)$}{v3,o1}
\fmf{phantom}{w3,o2}
\end{fmfgraph*}
\end{fmffile}
\end{minipage}
\begin{minipage}[c]{8.5cm}
\[
\begin{array}{c}
-\frac{e e_{{\rm eff}}^2}{4(m^2-M^2)(p^2\Lambda^{-2}-1)^2} \int 
\frac{d^4 k}{(2\pi)^4} \\
\frac{
\bar{u}(q_2) 
(\gamma^\nu \slashed{k} - \slashed{k} \gamma^\nu)
(\slashed{q_2} - \slashed{k} + M)
\gamma_\nu
(\slashed{q_2}+M)
(\gamma^\mu \slashed{p} - \slashed{p} \gamma^\mu)
u(q_1)
}
{
((k-q_2)^2 - M^2 + i\epsilon)
(k^2+i\epsilon)
(k^2\Lambda^{-2} - 1)^2
}
\end{array}
\]
\end{minipage}

\medskip

\noindent
3'.
\begin{minipage}[t]{7cm}
\begin{fmffile}{muonmagmoment3prime}
\unitlength = 1mm
\begin{fmfgraph*}(70,20)
\fmfleft{i1,i2}
\fmfright{o1,o2}
\fmfv{decor.shape=circle,decor.filled=full,decor.size=2thick}{v1}
\fmfv{}{v2}
\fmfv{decor.shape=circle,decor.filled=full,decor.size=2thick}{v3}
\fmf{fermion,label=$\ell(q_1)$}{i1,v1}
\fmf{phantom}{i2,w1}
\fmf{fermion,label=$\ell^*$,label.side=right}{v1,v2}
\fmf{photon,left=0.8,tension=0,label=$k$}{v1,v2}
\fmf{phantom}{w1,w2}
\fmf{fermion,label=$\ell^*$,label.side=right}{v2,v3}
\fmf{phantom}{w2,w3}
\fmf{photon,tension=0,label=$p$}{w3,v3}
\fmf{fermion,label=$\ell(q_2)$}{v3,o1}
\fmf{phantom}{w3,o2}
\end{fmfgraph*}
\end{fmffile}
\end{minipage}
\begin{minipage}[c]{8.5cm}
\[
\begin{array}{c}
-\frac{e e_{{\rm eff}}^2}{4(m^2-M^2)(p^2\Lambda^{-2}-1)^2} \int 
\frac{d^4 k}{(2\pi)^4} \\
\frac{
\bar{u}(q_2) 
(\gamma^\mu \slashed{p} - \slashed{p} \gamma^\mu)
(\slashed{q_1}+M)
\gamma^\nu
(\slashed{q_1} - \slashed{k} + M)
(\gamma_\nu \slashed{k} - \slashed{k} \gamma_\nu)
u(q_1)
}
{
((k-q_1)^2 - M^2 + i\epsilon)
(k^2+i\epsilon)
(k^2\Lambda^{-2} - 1)^2
}
\end{array}
\]
\end{minipage}

\medskip

\noindent
4.
\begin{minipage}[t]{7cm}
\begin{fmffile}{muonmagmoment4}
\unitlength = 1mm
\begin{fmfgraph*}(70,20)
\fmfleft{i1,i2}
\fmfright{o1,o2}
\fmfv{decor.shape=circle,decor.filled=full,decor.size=2thick}{v1}
\fmfv{decor.shape=circle,decor.filled=full,decor.size=2thick}{v2}
\fmfv{}{v3}
\fmf{fermion,label=$\ell(q_1)$}{i1,v1}
\fmf{phantom}{i2,w1}
\fmf{photon,tension=0,label=$p$}{w1,v1}
\fmf{fermion,label=$\ell^*$,label.side=left}{v1,v2}
\fmf{phantom}{w1,w2}
\fmf{fermion,label=$\ell$,label.side=right}{v2,v3}
\fmf{phantom}{w2,w3}
\fmf{photon,left=0.8,tension=0,label=$k$}{v2,v3}
\fmf{fermion,label=$\ell(q_2)$}{v3,o1}
\fmf{phantom}{w3,o2}
\end{fmfgraph*}
\end{fmffile}
\end{minipage}
\begin{minipage}[c]{8.5cm}
\[
\begin{array}{c}
-\frac{e e_{{\rm eff}}^2}{4(m^2-M^2)(p^2\Lambda^{-2}-1)^2} \int 
\frac{d^4 k}{(2\pi)^4} \\
\frac{
\bar{u}(q_2) 
\gamma^\nu
(\slashed{q_2} - \slashed{k} + M)
(\gamma_\nu \slashed{k} - \slashed{k} \gamma_\nu)
(\slashed{q_2}+M)
(\gamma^\mu \slashed{p} - \slashed{p} \gamma^\mu)
u(q_1)
}
{
((k-q_2)^2 - M^2 + i\epsilon)
(k^2+i\epsilon)
(k^2\Lambda^{-2} - 1)^2
}
\end{array}
\]
\end{minipage}

\medskip

\noindent
4'.
\begin{minipage}[t]{7cm}
\begin{fmffile}{muonmagmoment4prime}
\unitlength = 1mm
\begin{fmfgraph*}(70,20)
\fmfleft{i1,i2}
\fmfright{o1,o2}
\fmfv{}{v1}
\fmfv{decor.shape=circle,decor.filled=full,decor.size=2thick}{v2}
\fmfv{decor.shape=circle,decor.filled=full,decor.size=2thick}{v3}
\fmf{fermion,label=$\ell(q_1)$}{i1,v1}
\fmf{phantom}{i2,w1}
\fmf{fermion,label=$\ell$,label.side=right}{v1,v2}
\fmf{photon,left=0.8,tension=0,label=$k$}{v1,v2}
\fmf{phantom}{w1,w2}
\fmf{fermion,label=$\ell^*$,label.side=right}{v2,v3}
\fmf{phantom}{w2,w3}
\fmf{photon,tension=0,label=$p$}{w3,v3}
\fmf{fermion,label=$\ell(q_2)$}{v3,o1}
\fmf{phantom}{w3,o2}
\end{fmfgraph*}
\end{fmffile}
\end{minipage}
\begin{minipage}[c]{8.5cm}
\[
\begin{array}{c}
-\frac{e e_{{\rm eff}}^2}{4(m^2-M^2)(p^2\Lambda^{-2}-1)^2} \int 
\frac{d^4 k}{(2\pi)^4} \\
\frac{
\bar{u}(q_2) 
(\gamma^\mu \slashed{p} - \slashed{p} \gamma^\mu)
(\slashed{q_1}+M)
(\gamma^\nu \slashed{k} - \slashed{k} \gamma^\nu)
(\slashed{q_1} - \slashed{k} + M)
\gamma_\nu
u(q_1)
}
{
((k-q_1)^2 - M^2 + i\epsilon)
(k^2+i\epsilon)
(k^2\Lambda^{-2} - 1)^2
}
\end{array}
\]
\end{minipage}
\caption{
\label{table:matrixelement}
Feynman graphs and their matrix element contribution to muon magnetic
moment.
}
\end{table}

\noindent
\begin{table}
\begin{minipage}[t]{7cm}
\begin{center}
Diagram
\end{center}
\end{minipage}
\begin{minipage}[b]{8.5cm}
\begin{center}
$F_2(0)$
\end{center}
\end{minipage}

\medskip

\noindent
1.
\begin{minipage}[t]{7cm}
\begin{fmffile}{muonmagmoment1}
\unitlength = 1mm
\begin{fmfgraph*}(70,20)
\fmfleft{i1,i2}
\fmfright{o1,o2}
\fmfv{decor.shape=circle,decor.filled=full,decor.size=2thick}{v1}
\fmfv{decor.shape=circle,decor.filled=full,decor.size=2thick}{v3}
\fmfv{}{v2}
\fmf{fermion,label=$\ell(q_1)$}{i1,v1}
\fmf{fermion,label=$\ell^*$,label.side=left}{v1,v2}
\fmf{fermion,label=$\ell^*$,label.side=left}{v2,v3}
\fmf{fermion,label=$\ell(q_2)$}{v3,o1}
\fmf{phantom}{i2,v4,o2}
\fmf{photon,tension=0,label=$p$}{v4,v2}
\fmf{photon,right=0.5,tension=0,label=$k$}{v1,v3}
\end{fmfgraph*}
\end{fmffile}
\end{minipage}
\begin{minipage}[c]{8.5cm}
\[
(f_1 + f_2 + f'')^2 \cdot \frac{\alpha}{2\pi} \cdot \frac{8Mm}{\Lambda^2} \cdot
I_1(m,M,\Lambda)
+ \mbox{low. ord.}
\]
\end{minipage}

\medskip

\[
I_1(m,M,\Lambda) =
\int_0^1 \int_0^{1-x} \int_0^{1-x-y} dx dy dz \,
\frac{
(1-y-z)(1-x-y-z)^3
}
{
[-(y+z)(1-y-z)m^2\Lambda^{-2} + (y+z)M^2\Lambda^{-2} + (1-x-y-z)]^4
}.
\]

\caption{
\label{table:anomalous1}
Feynman graph 1 and its contribution to the form factor $F_2(0)$.
}
\end{table}

\noindent
\begin{table}
\begin{minipage}[t]{7cm}
\begin{center}
Diagram
\end{center}
\end{minipage}
\begin{minipage}[b]{8.5cm}
\begin{center}
$F_2(0)$
\end{center}
\end{minipage}

\medskip

\noindent
2.
\begin{minipage}[t]{7cm}
\begin{fmffile}{muonmagmoment2}
\unitlength = 1mm
\begin{fmfgraph*}(70,20)
\fmfleft{i1,i2}
\fmfright{o1,o2}
\fmfv{decor.shape=circle,decor.filled=full,decor.size=2thick}{v1}
\fmfv{decor.shape=circle,decor.filled=full,decor.size=2thick}{v2}
\fmfv{}{v2}
\fmf{fermion,label=$\ell(q_1)$}{i1,v1}
\fmf{fermion,label=$\ell^*$,label.side=left}{v1,v2}
\fmf{fermion,label=$\ell$,label.side=left}{v2,v3}
\fmf{fermion,label=$\ell(q_2)$}{v3,o1}
\fmf{phantom}{i2,v4,o2}
\fmf{photon,tension=0,label=$p$}{v4,v2}
\fmf{photon,right=0.5,tension=0,label=$k$}{v1,v3}
\end{fmfgraph*}
\end{fmffile}
\end{minipage}
\begin{minipage}[b]{8.5cm}
\[
-(f_1 + f_2 + f'')^2 \cdot \frac{\alpha}{2\pi} \cdot \frac{8Mm}{\Lambda^2} \cdot
I_2(m,M,\Lambda) - \mbox{low. ord.}
\]
\end{minipage}

\medskip

\noindent
2'.
\begin{minipage}[t]{7cm}
\begin{fmffile}{muonmagmoment2prime}
\unitlength = 1mm
\begin{fmfgraph*}(70,20)
\fmfleft{i1,i2}
\fmfright{o1,o2}
\fmfv{decor.shape=circle,decor.filled=full,decor.size=2thick}{v2}
\fmfv{decor.shape=circle,decor.filled=full,decor.size=2thick}{v3}
\fmfv{}{v2}
\fmf{fermion,label=$\ell(q_1)$}{i1,v1}
\fmf{fermion,label=$\ell^*$,label.side=left}{v1,v2}
\fmf{fermion,label=$\ell$,label.side=left}{v2,v3}
\fmf{fermion,label=$\ell(q_2)$}{v3,o1}
\fmf{phantom}{i2,v4,o2}
\fmf{photon,tension=0,label=$p$}{v4,v2}
\fmf{photon,right=0.5,tension=0,label=$k$}{v1,v3}
\end{fmfgraph*}
\end{fmffile}
\end{minipage}
\begin{minipage}[b]{8.5cm}
\[
(f_1 + f_2 + f'')^2 \cdot \frac{\alpha}{2\pi} \cdot \frac{8Mm}{\Lambda^2} \cdot
I_2(m,M,\Lambda) + \mbox{low. ord.}
\]
\end{minipage}

\medskip

\[
I_2(m,M,\Lambda) =
\int_0^1 \int_0^{1-x} \int_0^{1-x-y} dx dy dz \,
\frac{
1-x-y-z
}
{
[-(y+z)(1-y-z)m^2\Lambda^{-2} + (y+z)M^2\Lambda^{-2} + (1-x-y-z)]^2
}.
\]
\caption{
\label{table:anomalous2}
Feynman graphs 2, 2' and their contribution to the form factor $F_2(0)$.
}
\end{table}

\noindent
\begin{table}
\begin{minipage}[t]{7cm}
\begin{center}
Diagram
\end{center}
\end{minipage}
\begin{minipage}[b]{8.5cm}
\begin{center}
$F_2(0)$
\end{center}
\end{minipage}

\medskip

\noindent
3.
\begin{minipage}[t]{7cm}
\begin{fmffile}{muonmagmoment3}
\unitlength = 1mm
\begin{fmfgraph*}(70,20)
\fmfleft{i1,i2}
\fmfright{o1,o2}
\fmfv{decor.shape=circle,decor.filled=full,decor.size=2thick}{v1}
\fmfv{}{v2}
\fmfv{decor.shape=circle,decor.filled=full,decor.size=2thick}{v3}
\fmf{fermion,label=$\ell(q_1)$}{i1,v1}
\fmf{phantom}{i2,w1}
\fmf{photon,tension=0,label=$p$}{w1,v1}
\fmf{fermion,label=$\ell^*$,label.side=right}{v1,v2}
\fmf{phantom}{w1,w2}
\fmf{photon,left=0.8,tension=0,label=$k$}{v2,v3}
\fmf{fermion,label=$\ell^*$,label.side=right}{v2,v3}
\fmf{phantom}{w2,w3}
\fmf{fermion,label=$\ell(q_2)$}{v3,o1}
\fmf{phantom}{w3,o2}
\end{fmfgraph*}
\end{fmffile}
\end{minipage}
\begin{minipage}[b]{8.5cm}
\[
(f_1+f_2+f'')^2 \cdot \frac{\alpha}{2\pi} \cdot \frac{18m}{M-m} 
\cdot I_3(m,M,\Lambda)
\]
\end{minipage}

\medskip

\noindent
3'.
\begin{minipage}[t]{7cm}
\begin{fmffile}{muonmagmoment3prime}
\unitlength = 1mm
\begin{fmfgraph*}(70,20)
\fmfleft{i1,i2}
\fmfright{o1,o2}
\fmfv{decor.shape=circle,decor.filled=full,decor.size=2thick}{v1}
\fmfv{}{v2}
\fmfv{decor.shape=circle,decor.filled=full,decor.size=2thick}{v3}
\fmf{fermion,label=$\ell(q_1)$}{i1,v1}
\fmf{phantom}{i2,w1}
\fmf{fermion,label=$\ell^*$,label.side=right}{v1,v2}
\fmf{photon,left=0.8,tension=0,label=$k$}{v1,v2}
\fmf{phantom}{w1,w2}
\fmf{fermion,label=$\ell^*$,label.side=right}{v2,v3}
\fmf{phantom}{w2,w3}
\fmf{photon,tension=0,label=$p$}{w3,v3}
\fmf{fermion,label=$\ell(q_2)$}{v3,o1}
\fmf{phantom}{w3,o2}
\end{fmfgraph*}
\end{fmffile}
\end{minipage}
\begin{minipage}[b]{8.5cm}
\[
-(f_1+f_2+f'')^2 \cdot \frac{\alpha}{2\pi} \cdot \frac{18m}{M-m} 
\cdot I_3(m,M,\Lambda)
\]
\end{minipage}

\medskip

\begin{eqnarray*}
I_3(m,M,\Lambda) 
& = &
\int_0^1 \int_0^{1-x} \int_0^{1-x-y} dx dy dz \,
\left(
\frac{x(1-x)m^2 - xMm}
{6 \Lambda^2 (x M^2\Lambda^{-2} - m^2\Lambda^{-2} x(1-x) + (1-x-y))^2} 
\right. \\
&   &
~~~~~~~~~~~~~~~~~~~~~~~~~~~~~~~~
\left.
{} + 
\frac{1}{3(x M^2\Lambda^{-2} - m^2\Lambda^{-2} x(1-x) + (1-x-y))}
\right)
\end{eqnarray*}

\caption{
\label{table:anomalous3}
Feynman graphs 3, 3' and their contribution to the form factor $F_2(0)$.
}
\end{table}

\noindent
\begin{table}
\begin{minipage}[t]{7cm}
\begin{center}
Diagram
\end{center}
\end{minipage}
\begin{minipage}[b]{8.5cm}
\begin{center}
$F_2(0)$
\end{center}
\end{minipage}

\medskip

\noindent
4.
\begin{minipage}[t]{7cm}
\begin{fmffile}{muonmagmoment4}
\unitlength = 1mm
\begin{fmfgraph*}(70,20)
\fmfleft{i1,i2}
\fmfright{o1,o2}
\fmfv{decor.shape=circle,decor.filled=full,decor.size=2thick}{v1}
\fmfv{decor.shape=circle,decor.filled=full,decor.size=2thick}{v2}
\fmfv{}{v3}
\fmf{fermion,label=$\ell(q_1)$}{i1,v1}
\fmf{phantom}{i2,w1}
\fmf{photon,tension=0,label=$p$}{w1,v1}
\fmf{fermion,label=$\ell^*$,label.side=left}{v1,v2}
\fmf{phantom}{w1,w2}
\fmf{fermion,label=$\ell$,label.side=right}{v2,v3}
\fmf{phantom}{w2,w3}
\fmf{photon,left=0.8,tension=0,label=$k$}{v2,v3}
\fmf{fermion,label=$\ell(q_2)$}{v3,o1}
\fmf{phantom}{w3,o2}
\end{fmfgraph*}
\end{fmffile}
\end{minipage}
\begin{minipage}[b]{8.5cm}
\[
(f_1+f_2+f'')^2 \cdot \frac{\alpha}{2\pi} \cdot \frac{18m}{M-m} 
\cdot I_4(m,\Lambda)
\]
\end{minipage}

\medskip

\noindent
4'.
\begin{minipage}[t]{7cm}
\begin{fmffile}{muonmagmoment4prime}
\unitlength = 1mm
\begin{fmfgraph*}(70,20)
\fmfleft{i1,i2}
\fmfright{o1,o2}
\fmfv{}{v1}
\fmfv{decor.shape=circle,decor.filled=full,decor.size=2thick}{v2}
\fmfv{decor.shape=circle,decor.filled=full,decor.size=2thick}{v3}
\fmf{fermion,label=$\ell(q_1)$}{i1,v1}
\fmf{phantom}{i2,w1}
\fmf{fermion,label=$\ell$,label.side=right}{v1,v2}
\fmf{photon,left=0.8,tension=0,label=$k$}{v1,v2}
\fmf{phantom}{w1,w2}
\fmf{fermion,label=$\ell^*$,label.side=right}{v2,v3}
\fmf{phantom}{w2,w3}
\fmf{photon,tension=0,label=$p$}{w3,v3}
\fmf{fermion,label=$\ell(q_2)$}{v3,o1}
\fmf{phantom}{w3,o2}
\end{fmfgraph*}
\end{fmffile}
\end{minipage}
\begin{minipage}[c]{8.5cm}
\[
-(f_1+f_2+f'')^2 \cdot \frac{\alpha}{2\pi} \cdot \frac{18m}{M-m} 
\cdot I_4(m,\Lambda)
\]
\end{minipage}

\medskip

\[
I_4(m,\Lambda) =
\int_0^1 \int_0^{1-x} \int_0^{1-x-y} dx dy dz \,
\left(
\frac{m^2 x^2}
{6\Lambda^2 (x^2 m^2\Lambda^{-2}  + (1-x-y))^2} - 
\frac{1}
{3(x^2 m^2\Lambda^{-2}  + (1-x-y))}
\right)
\]

\caption{
\label{table:anomalous4}
Feynman graphs 4, 4' and their contribution to the form factor $F_2(0)$.
}
\end{table}

\appendix

\section{Details of the calculation for Diagram 1}
\label{app:calculation}
We now provide some details of the calculation of $F_2(0)$ for Diagram~1.
The Feynman integral is
\begin{eqnarray*}
\lefteqn{
i M^\mu 
} \\
& = &
\int \frac{d^4 k}{(2\pi)^4} \, 
\frac{-i g^{\nu \alpha}}{k^2+i\epsilon} \\
&   &
~~~~~~~~
\bar{u}(q_2) 
(e_{{\rm eff}} k_\beta \sigma^{\nu\beta})
\frac{1}{(1-\frac{k^2}{\Lambda^2})^2}
\frac{
i(\slashed{q_2} + \slashed{k} + M)
}
{
(q_2+k)^2 - M^2 + i\epsilon
}
(-i e \gamma^\mu) \\
&   &
~~~~~~~~~~~~~~~~~~
\frac{
i (\slashed{q_1} + \slashed{k} + M)
}
{
(q_1+k)^2 - M^2 + i\epsilon
}
(e_{{\rm eff}} k_\beta\sigma^{\alpha\beta})
\frac{1}{(1-\frac{k^2}{\Lambda^2})^2} 
u(q_1) \\
& = &
-\frac{e e_{{\rm eff}}^2}{4} \int 
\frac{d^4 k}{(2\pi)^4} \, \frac{
\bar{u}(q_2) 
(\gamma^\nu \slashed{k} - \slashed{k} \gamma^\nu)
(\slashed{q_2} + \slashed{k} + M)
\gamma^\mu
(\slashed{q_1} + \slashed{k} + M)
(\gamma_\nu \slashed{k} - \slashed{k} \gamma_\nu)
u(q_1)
}
{
A_1 A_2 A_3 A'_4 A'_5 A'_6 A'_7
},
\end{eqnarray*}
where
\begin{eqnarray*} 
A_1
& = &
k^2 + i\epsilon, \\
A_2
& = &
(k + q_1)^2 - M^2 + i\epsilon, \\
A_3
& = &
(k + q_2)^2 - M^2 + i\epsilon, \\
A_4
& = &
k^2 - \Lambda^2, ~~~~~
A'_4\;=\;\Lambda^{-2} A_4, \\
A_5
& = &
A_6 \;=\;A_7\;=\;A_4, ~~~~~
A'_5\;=\;A'_6\;=\;A'_7\;=\;A'_4.
\end{eqnarray*} 

Using standard results about contractions with Dirac matrices, we can simplify
the integral to get
\begin{eqnarray*}
i M^\mu 
& = &
-\frac{e e_{{\rm eff}}^2 \Lambda^8}{4} \int 
\frac{d^4 k}{(2\pi)^4} \, 
\frac{1}{A_1 A_2 A_3 A_4 A_5 A_6 A_7} \\
&   &
~~~~~~~~~~~~~~~~~
\bar{u}(q_2) 
[
4 \slashed{k} (\slashed{q_1} + \slashed{k})
\gamma^\mu (\slashed{q_2} + \slashed{k}) \slashed{k} \\
&   &
~~~~~~~~~~~~~~~~~~~~~~~~~
{} - 
8 M (q_1^\mu + k^\mu) \slashed{k} \slashed{k} \\
&   &
~~~~~~~~~~~~~~~~~~~~~~~~~
{} - 
8 M (q_2^\mu + k^\mu) \slashed{k} \slashed{k} \\
&   &
~~~~~~~~~~~~~~~~~~~~~~~~
{} + 
8 M^2 k^\mu \slashed{k} \\
&   &
~~~~~~~~~~~~~~~~~~~~~~~~~
{} +
4 \slashed{k}
(\slashed{q_2} + \slashed{k}) \slashed{k} (\slashed{q_1} + \slashed{k}) \gamma^\mu 
+ 4 \slashed{k} \gamma^\mu (\slashed{q_1} + \slashed{k}) \slashed{k} 
(\slashed{q_2} + \slashed{k}) \\
&   &
~~~~~~~~~~~~~~~~~~~~~~~~
{} - 
4 M \slashed{k} \slashed{k} (\slashed{q_1} + \slashed{k}) \gamma^\mu \\
&   &
~~~~~~~~~~~~~~~~~~~~~~~~
{} -
4 M \slashed{k} \slashed{k} \gamma^\mu (\slashed{q_2} + \slashed{k}) \\
&   &
~~~~~~~~~~~~~~~~~~~~~~~~
{} + 
4 M^2 \slashed{k} \gamma^\mu \slashed{k}
]
u(q_1).
\end{eqnarray*} 

In order to evaluate the above expression, we will complete the square in the
denominator. Define 
$
D = x A_1 + y A_2 + z A_3 + w_4 A_4 + w_5 A_5 + w_6 A_6 + w_7 A_7,
$
where $x, y, z, w_4, \ldots, w_7 \geq 0$, $x+y+z+w_4+\cdots+w_7=1$.
Completing the square using standard techniques gives us
\begin{eqnarray*}
D 
& = &
(k+ (y + z) q_1 + z p)^2 - \Delta + (x+y+z)i\epsilon, 
\end{eqnarray*}
where 
$
\Delta = -(y + z)(1 - y - z)m^2 - y z p^2 + (y + z) M^2 + (1 - x - y - z) \Lambda^2.
$ 
This leads to the following expression for $i M^\mu$.
\begin{eqnarray*}
\lefteqn{
i M^\mu 
} \\
& = &
-\frac{6! e e_{{\rm eff}}^2 \Lambda^8}{4} 
\int \frac{d^4 k}{(2\pi)^4} \, 
\int_0^1 dx dy dz dw_4 \cdots dw_7 \, \delta(x+y+z+w_4+\cdots+w_7-1) \\
&   &
~~~~~~~~~~~~~~~~~~~~~~~~~~~~~~~
\frac{1}
{
[(k + (y + z) q_1 + z p)^2-\Delta+(x+y+z)i\epsilon]^7
} \\
&   &
~~~~~~~~~~~~~~~~~~~~~~~~~~~~~~~
\bar{u}(q_2) 
[
4 \slashed{k} (\slashed{q_1} + \slashed{k})
\gamma^\mu (\slashed{q_2} + \slashed{k}) \slashed{k} \\
&   &
~~~~~~~~~~~~~~~~~~~~~~~~~~~~~~~~~~~~~~~
{} - 
8 M (q_1^\mu + k^\mu) \slashed{k} \slashed{k} \\
&   &
~~~~~~~~~~~~~~~~~~~~~~~~~~~~~~~~~~~~~~~
{} - 
8 M (q_2^\mu + k^\mu) \slashed{k} \slashed{k} \\
&   &
~~~~~~~~~~~~~~~~~~~~~~~~~~~~~~~~~~~~~~~
{} + 
8 M^2 k^\mu \slashed{k} \\
&   &
~~~~~~~~~~~~~~~~~~~~~~~~~~~~~~~~~~~~~~~
{} +
4 \slashed{k}
(\slashed{q_2} + \slashed{k}) \slashed{k} (\slashed{q_1} + \slashed{k}) \gamma^\mu 
+ 4 \slashed{k} \gamma^\mu (\slashed{q_1} + \slashed{k}) \slashed{k} 
(\slashed{q_2} + \slashed{k}) \\
&   &
~~~~~~~~~~~~~~~~~~~~~~~~~~~~~~~~~~~~~~
{} - 
4 M \slashed{k} \slashed{k} (\slashed{q_1} + \slashed{k}) \gamma^\mu \\
&   &
~~~~~~~~~~~~~~~~~~~~~~~~~~~~~~~~~~~~~~
{} -
4 M \slashed{k} \slashed{k} \gamma^\mu (\slashed{q_2} + \slashed{k}) \\
&   &
~~~~~~~~~~~~~~~~~~~~~~~~~~~~~~~~~~~~~~
{} + 
4 M^2 \slashed{k} \gamma^\mu \slashed{k}
]
u(q_1).
\end{eqnarray*}

Performing the change of variable 
$k^\mu \mapsto k^\mu - (y + z) q_1^\mu - z p^\mu = k^\mu - y q_1^\mu - z q_2^\mu$ 
leads to
\begin{eqnarray*}
i M^\mu 
& = &
-\frac{6! e e_{{\rm eff}}^2 \Lambda^8}{4} 
\int \frac{d^4 k}{(2\pi)^4} \, 
\int_0^1 dx dy dz dw_4 \cdots dw_7 \, \delta(x+y+z+w_4+\cdots+w_7-1) \\
&   &
~~~~~~~~~~~~~~~~~~~~~~~~~~~~
\frac{
N'^\mu
}
{
[k^2-\Delta+(x+y+z)i\epsilon]^7
},
\end{eqnarray*}
where
\begin{eqnarray*}
N'^\mu 
& = &
\bar{u}(q_2) 
[
4 (\slashed{k} - y \slashed{q_1} - z \slashed{q_2})
(\slashed{k} + (1 - y) \slashed{q_1} - z \slashed{q_2})
\gamma^\mu 
(\slashed{k} - y \slashed{q_1} + (1 - z) \slashed{q_2}) 
(\slashed{k} - y \slashed{q_1} - z \slashed{q_2}) \\
&   &
~~~~~~~~
{} - 
8 M (k^\mu + (1 - y) q_1^\mu - z q_2^\mu) 
(\slashed{k} - y \slashed{q_1} - z \slashed{q_2}) 
(\slashed{k} - y \slashed{q_1} - z \slashed{q_2}) \\
&   &
~~~~~~~~
{} - 
8 M (k^\mu - y q_1^\mu + (1 - z) q_2^\mu) 
(\slashed{k} - y \slashed{q_1} - z \slashed{q_2}) 
(\slashed{k} - y \slashed{q_1} - z \slashed{q_2}) \\
&   &
~~~~~~~~
{} + 
8 M^2 (k^\mu  - y q_1^\mu - z q_2^\mu)
(\slashed{k} - y \slashed{q_1} - z \slashed{q_2}) \\
&   &
~~~~~~~~
{} +
4 (\slashed{k} - y \slashed{q_1} - z \slashed{q_2}) 
(\slashed{k} - y \slashed{q_1} + (1 - z) \slashed{q_2}) 
(\slashed{k} - y \slashed{q_1} - z \slashed{q_2}) 
(\slashed{k} + (1 - y) \slashed{q_1} - z \slashed{q_2})
\gamma^\mu \\
&   &
~~~~~~~~
{} +
4 (\slashed{k} - y \slashed{q_1} - z \slashed{q_2}) 
\gamma^\mu 
(\slashed{k} + (1 - y) \slashed{q_1} - z \slashed{q_2})
(\slashed{k} - y \slashed{q_1} - z \slashed{q_2}) 
(\slashed{k} - y \slashed{q_1} + (1 - z) \slashed{q_2}) \\
&   &
~~~~~~~~
{} - 
4 M 
(\slashed{k} - y \slashed{q_1} - z \slashed{q_2}) 
(\slashed{k} - y \slashed{q_1} - z \slashed{q_2}) 
(\slashed{k} + (1 - y) \slashed{q_1} - z \slashed{q_2})
\gamma^\mu \\
&   &
~~~~~~~~
{} -
4 M 
(\slashed{k} - y \slashed{q_1} - z \slashed{q_2}) 
(\slashed{k} - y \slashed{q_1} - z \slashed{q_2}) 
\gamma^\mu 
(\slashed{k} - y \slashed{q_1} + (1 - z) \slashed{q_2}) \\
&   &
~~~~~~~~
{} + 
4 M^2 
(\slashed{k} - y \slashed{q_1} - z \slashed{q_2}) 
\gamma^\mu 
(\slashed{k} - y \slashed{q_1} - z \slashed{q_2}) 
]
u(q_1).
\end{eqnarray*}

Using standard techniques of evaluating Feynman integrals and a fair amount of
simplification, it suffices to consider
the integral of an expression with the following as numerator instead of $N'^\mu$:
\begin{align*}
N^\mu 
& = 
\bar{u}(q_2) 
[
4 (k^2)^2 \gamma^\mu ~~~~~~ I  \\ 
&   
~~~~~~~~~~~~~
{} +
4 k^2
\gamma^\mu 
(- y \slashed{q_1} + (1 - z) \slashed{q_2}) 
(- y \slashed{q_1} - z \slashed{q_2}) ~~~~~~ II  \\ 
&   
~~~~~~~~~~~~~
{} +
4 k^2
((1 - y) q_1^\mu - z q_2^\mu)
(- y \slashed{q_1} - z \slashed{q_2}) ~~~~~~ II  \\ 
&   
~~~~~~~~~~~~~
{} -
2 k^2
(- y \slashed{q_1} + (1 - z) \slashed{q_2}) 
\gamma^\mu 
((1 - y) \slashed{q_1} - z \slashed{q_2}) ~~~~~~ II  \\ 
&   
~~~~~~~~~~~~~
{} -
2 k^2 
(- y \slashed{q_1} - z \slashed{q_2})
\gamma^\mu 
(- y \slashed{q_1} - z \slashed{q_2}) ~~~~~~ II  \\ 
&   
~~~~~~~~~~~~~
{} +
4 k^2 
(- y q_1^\mu + (1 - z) q_2^\mu) 
(- y \slashed{q_1} - z \slashed{q_2}) ~~~~~~ II  \\ 
&   
~~~~~~~~~~~~~
{} +
4 k^2 (- y \slashed{q_1} - z \slashed{q_2})
((1 - y) \slashed{q_1} - z \slashed{q_2})
\gamma^\mu ~~~~~~ II  \\ 
&   
~~~~~~~~~~~~~
{} +
4 (- y \slashed{q_1} - z \slashed{q_2})
((1 - y) \slashed{q_1} - z \slashed{q_2})
\gamma^\mu 
(- y \slashed{q_1} + (1 - z) \slashed{q_2}) 
(- y \slashed{q_1} - z \slashed{q_2}) ~~~~~~ II  \\ 
&   
~~~~~~~~~~~~~
{} +
4 (k^2)^2
\gamma^\mu ~~~~~~ I  \\ 
&   
~~~~~~~~~~~~~
{} +
4 k^2
(- y \slashed{q_1} - z \slashed{q_2}) 
((1 - y) \slashed{q_1} - z \slashed{q_2})
\gamma^\mu ~~~~~~ II  \\ 
&   
~~~~~~~~~~~~~
{} -
2 k^2
(- y \slashed{q_1} + (1 - z) \slashed{q_2}) 
((1 - y) \slashed{q_1} - z \slashed{q_2})
\gamma^\mu ~~~~~~ II  \\ 
&   
~~~~~~~~~~~~~
{} +
4 k^2
((- y q_1 + (1 - z) q_2) \cdot (- y q_1 - z q_2))
\gamma^\mu ~~~~~~ II  \\ 
&   
~~~~~~~~~~~~~
{} +
4 k^2 
(- y \slashed{q_1} - z \slashed{q_2}) 
((1 - y) \slashed{q_1} - z \slashed{q_2})
\gamma^\mu ~~~~~~ II  \\ 
&   
~~~~~~~~~~~~~
{} -
2 k^2 (- y \slashed{q_1} - z \slashed{q_2}) 
(- y \slashed{q_1} - z \slashed{q_2}) 
\gamma^\mu ~~~~~~ II  \\ 
&   
~~~~~~~~~~~~~
{} +
4 k^2 (- y \slashed{q_1} - z \slashed{q_2}) 
(- y \slashed{q_1} + (1 - z) \slashed{q_2}) 
\gamma^\mu ~~~~~~ II  \\ 
&   
~~~~~~~~~~~~~
{} +
4 (- y \slashed{q_1} - z \slashed{q_2}) 
(- y \slashed{q_1} + (1 - z) \slashed{q_2}) 
(- y \slashed{q_1} - z \slashed{q_2}) 
((1 - y) \slashed{q_1} - z \slashed{q_2})
\gamma^\mu ~~~~~~ II  \\ 
&   
~~~~~~~~~~~~~
{} -
2 (k^2)^2
\gamma^\mu ~~~~~~ I  \\ 
&   
~~~~~~~~~~~~~
{} -
2 k^2 
\gamma^\mu 
(- y \slashed{q_1} - z \slashed{q_2}) 
(- y \slashed{q_1} + (1 - z) \slashed{q_2}) ~~~~~~ II  \\ 
&   
~~~~~~~~~~~~~
{} +
4 k^2 
((1 - y) q_1^\mu - z q_2^\mu)
(- y \slashed{q_1} + (1 - z) \slashed{q_2}) ~~~~~~ II  \\ 
&   
~~~~~~~~~~~~~
{} -
2 k^2
(- y \slashed{q_1} - z \slashed{q_2}) 
((1 - y) \slashed{q_1} - z \slashed{q_2})
\gamma^\mu ~~~~~~ II  \\ 
&   
~~~~~~~~~~~~~
{} +
4 k^2 
(- y \slashed{q_1} - z \slashed{q_2}) 
\gamma^\mu 
(- y \slashed{q_1} + (1 - z) \slashed{q_2}) ~~~~~~ II  \\ 
&   
~~~~~~~~~~~~~
{} -
2 k^2 (- y \slashed{q_1} - z \slashed{q_2}) 
\gamma^\mu 
(- y \slashed{q_1} - z \slashed{q_2}) ~~~~~~ II  \\ 
&   
~~~~~~~~~~~~~
{} +
4 k^2 (- y \slashed{q_1} - z \slashed{q_2}) 
\gamma^\mu 
((1 - y) \slashed{q_1} - z \slashed{q_2}) ~~~~~~ II  \\ 
&   
~~~~~~~~~~~~~
{} +
4 (- y \slashed{q_1} - z \slashed{q_2}) 
\gamma^\mu 
((1 - y) \slashed{q_1} - z \slashed{q_2})
(- y \slashed{q_1} - z \slashed{q_2}) 
(- y \slashed{q_1} + (1 - z) \slashed{q_2}) ~~~~~~ II  \\ 
&   
~~~~~~~~~~~~~
{} - 
2 M k^2 \gamma^\mu 
(- y \slashed{q_1} - z \slashed{q_2}) ~~~~~~ III  \\ 
&   
~~~~~~~~~~~~~
{} - 
2 M k^2 
(- y \slashed{q_1} - z \slashed{q_2}) 
\gamma^\mu ~~~~~~ III  \\ 
&   
~~~~~~~~~~~~~
{} - 
8 M k^2 ((1 - 2y) q_1^\mu + (1 - 2z) q_2^\mu) ~~~~~~ III  \\
&   
~~~~~~~~~~~~~
{} - 
8 M ((1 - 2y) q_1^\mu + (1 - 2z) q_2^\mu) 
(- y \slashed{q_1} - z \slashed{q_2}) 
(- y \slashed{q_1} - z \slashed{q_2}) ~~~~~~ II  \\ 
&   
~~~~~~~~~~~~~
{} + 
2 M^2 k^2 \gamma^\mu ~~~~~~ I  \\ 
&   
~~~~~~~~~~~~~
{} + 
8 M^2 (- y q_1^\mu - z q_2^\mu)
(- y \slashed{q_1} - z \slashed{q_2}) ~~~~~~ II  \\ 
&   
~~~~~~~~~~~~~
{} - 
4 M k^2
((1 - y) \slashed{q_1} - z \slashed{q_2})
\gamma^\mu ~~~~~~ III  \\ 
&   
~~~~~~~~~~~~~
{} + 
2 M k^2
(- y \slashed{q_1} - z \slashed{q_2}) 
\gamma^\mu ~~~~~~ III  \\ 
&   
~~~~~~~~~~~~~
{} - 
4 M k^2
(- y \slashed{q_1} - z \slashed{q_2}) 
\gamma^\mu ~~~~~~ III  \\ 
&   
~~~~~~~~~~~~~
{} - 
4 M 
(- y \slashed{q_1} - z \slashed{q_2}) 
(- y \slashed{q_1} - z \slashed{q_2}) 
((1 - y) \slashed{q_1} - z \slashed{q_2})
\gamma^\mu ~~~~~~ II  \\ 
&   
~~~~~~~~~~~~~
{} -
4 M k^2
\gamma^\mu 
(- y \slashed{q_1} + (1 - z) \slashed{q_2}) ~~~~~~ III  \\ 
&   
~~~~~~~~~~~~~
{} -
4 M k^2
(- y q_1^\mu - z q_2^\mu) ~~~~~~ III  \\ 
&   
~~~~~~~~~~~~~
{} +
2 M k^2
(- y \slashed{q_1} - z \slashed{q_2}) 
\gamma^\mu  ~~~~~~ III  \\ 
&   
~~~~~~~~~~~~~
{} -
4 M 
(- y \slashed{q_1} - z \slashed{q_2}) 
(- y \slashed{q_1} - z \slashed{q_2}) 
\gamma^\mu 
(- y \slashed{q_1} + (1 - z) \slashed{q_2}) ~~~~~~ II  \\ 
&   
~~~~~~~~~~~~~
{} - 
2 M^2 k^2
\gamma^\mu  ~~~~~~ I  \\ 
&   
~~~~~~~~~~~~~
{} + 
4 M^2 
(- y \slashed{q_1} - z \slashed{q_2}) 
\gamma^\mu 
(- y \slashed{q_1} - z \slashed{q_2}) ~~~~~~ II  
]
u(q_1).
\end{align*}

The terms above marked I are all proportional to $\gamma^\mu$ and do not
contribute to the anomalous magnetic moment. The terms marked II do contribute,
but proportional to $k^2$ or $M^2$ or $M$ or $1$ in the numerator. Since the
denominator behaves like $(k^2 - \Delta)^7$, integration over the four momentum
$k$ will give contributions proportional to $\Delta^{-4}$, $M^2 \Delta^{-5}$,
$M \Delta^{-5}$ or $\Delta^{-5}$ respectively. Since $\Delta$ behaves like
$\Lambda^2$ times a bounded function of $(x,y,z)$, the contributions will be of
the order of $\Lambda^{-8}$, $M^2 \Lambda^{-10}$, $M \Lambda^{-10}$ and
$\Lambda^{-10}$ respectively. The terms marked III contribute to the anomalous
magnetic moment proportional to $M k^2$ in the numerator which, after integration
over $k$, end up being of the order of $M \Lambda^{-8}$. Thus,
in the regime $m \ll M \leq \Lambda$ and large $M$, the most significant 
contribution comes
from factors of order $M k^2$ multiplying $q_2^\mu$, $q_1^\mu$.  Collecting the
terms marked III we get
\begin{eqnarray*}
\hat{N}'^\mu
& = &
\bar{u}(q_2) 
[
- 2 M k^2 \gamma^\mu (-y\slashed{q_1} - z\slashed{q_2}) 
- 2 M k^2 (-y\slashed{q_1} - z\slashed{q_2}) \gamma^\mu 
- 8 M k^2 ((1 - 2y) q_1^\mu + (1 - 2z) q_2^\mu) \\
&   &
~~~~~~~~
{} - 
4 M k^2 ((1-y)\slashed{q_1} - z\slashed{q_2}) \gamma^\mu 
- 4 M k^2 \gamma^\mu (-y\slashed{q_1} + (1-z)\slashed{q_2}) 
- 4 M k^2 (-y q_1^\mu -z q_2^\mu) 
]
u(q_1) \\
& = &
\bar{u}(q_2) 
[
- 4 M k^2 (-y q_1^\mu - z q_2^\mu) 
- 8 M k^2 ((1 - 2y) q_1^\mu + (1 - 2z) q_2^\mu) \\
&   &
~~~~~~~~
{} + 
4 M k^2 (1-y) \gamma^\mu \slashed{q_1} - 8 M k^2 (1-y) q_1^\mu 
+ 4 M k^2 z \slashed{q_2} \gamma^\mu  
+ 4 M k^2 y \gamma^\mu \slashed{q_1} \\
&   &
~~~~~~~~
{} + 
 4 M k^2 (1-z) \slashed{q_2} \gamma^\mu 
- 8 M k^2 (1-z) q_2^\mu 
- 4 M k^2 (-y q_1^\mu -z q_2^\mu 
]
u(q_1) \\
& = &
\bar{u}(q_2) 
[
- 4 M k^2 ((4-8y) q_1^\mu + (4-8z) q_2^\mu) 
+ 4 M m k^2 (1-y) \gamma^\mu 
+ 4 M m k^2 z \gamma^\mu \\
&   &
~~~~~~~~
{} + 
4 M m k^2 y \gamma^\mu 
+ 4 M m k^2 (1-z) \gamma^\mu 
]
u(q_1) \\
& = &
\bar{u}(q_2) 
[
- 4 M k^2 ((4-8y) q_1^\mu + (4-8z) q_2^\mu) 
+ 8 M m k^2 \gamma^\mu 
]
u(q_1).
\end{eqnarray*}

The term above proportional to $\gamma^\mu$ does not contribute to the anomalous
magnetic moment. Thus, the main contribution comes from
\begin{eqnarray*}
\hat{N}^\mu
& = &
\bar{u}(q_2) 
[
- 4 M k^2 ((4 - 8y) q_1^\mu + (4 - 8z) q_2^\mu)
]
u(q_1).
\end{eqnarray*}
We now do the substitutions 
$2q_2^\mu = (q_2^\mu + q_1^\mu) + p^\mu$,
$2q_1^\mu = (q_2^\mu + q_1^\mu) - p^\mu$ and get
\[
\hat{N}^\mu =
\bar{u}(q_2) 
[
- 16 M k^2 
(
(1 - y - z) (q_2^\mu + q_1^\mu) +
(y - z) p^\mu
)
]
u(q_1).
\]
Applying the Gordon identity, we can simplify the above to get
\[
\hat{N}^\mu =
\bar{u}(q_2) 
[
- 16 M k^2 
(
(1 - y - z) (2m \gamma^\mu - i \sigma^{\mu\nu} p_\nu) +
(y - z) p^\mu
)
]
u(q_1).
\]
We can check that the $p^\mu$ term above goes to zero after
integrating over $x,y,z,w_4,\ldots,w_7$ because the integrand changes 
sign on swapping
$y$ and $z$. This satisfies the sanity check of the Ward identity. 
This implies that the only contribution to
the anomalous magnetic moment comes from the
$16i M k^2 (1-y-z) (\sigma^{\mu \nu} p_{\nu})$ term above.

Thus, the leading 
contribution to the anomalous magnet moment is captured by
$F_2(p^2)$ defined as follows:
\begin{eqnarray*}
F_2(p^2)
& = &
-\frac{2m \cdot 6! e e_{{\rm eff}}^2 \Lambda^8}{4e} \int 
\frac{d^4 k}{(2\pi)^4} \, 
\int_0^1 dx dy dz dw_4 \cdots dw_7 \, \delta(x+y+z+w_4+\cdots+w_7-1) \\
&   &
~~~~~~~~~~~~~~~~~~~~~~~~~~~~
\frac{
16i M k^2 (1-y-z)
}
{
[k^2-\Delta+(x+y+z)i\epsilon]^7
} +
\mbox{lower order},
\end{eqnarray*}
where 
$
\Delta = -(1-y-z)(y+z)m^2 - yzp^2 + (1-x-y-z)\Lambda^2 + (y+z)M^2. 
$ 
Using standard integration identities for Feynman integrals, we get
\begin{eqnarray*}
\lefteqn{
F_2(p^2)
} \\
& = &
-(8iMm \cdot 6! e^2 (f_1 + f_2 + f'')^2 \Lambda^6) 
\int_0^1 dx dy dz dw_4 \cdots dw_7 \, \delta(x+y+z+w_4+\cdots+w_7-1) \,
(1-y-z) \\
&   &
~~~~~~~~~~~~~~~~~~~~~~~~~~~~
\int \frac{d^4 k}{(2\pi)^4} \, 
\frac{k^2}{[k^2-\Delta+(x+y+z)i\epsilon]^7
} +
\mbox{lower order} \\
& = &
\frac{24 \alpha}{\pi} \cdot (f_1 + f_2 + f'')^2 \frac{Mm }{\Lambda^2}
\int_0^1 dx dy dz dw_4 \cdots dw_7 \, \delta(x+y+z+w_4+\cdots+w_7-1) \\
&   &
~~~~~~~~~~~~~~~~~~~~~~~~~~~~
\frac{
1-y-z
}
{
[-(1-y-z)(y+z)m^2\Lambda^{-2} - 
yz p^2 \Lambda^{-2} + (1-x-y-z) + (y+z)M^2\Lambda^{-2}
]^4 
} \\
&   &
{} + 
\mbox{lower order}.
\end{eqnarray*}

We only need to evaluate $F_2(0)$. This
gives us the further simplification that
\begin{eqnarray*}
F_2(0) 
& = &
\frac{24 \alpha}{\pi} \cdot \frac{Mm }{\Lambda^2} (f_1 + f_2 + f'')^2
\int_0^1 dx dy dz dw_4 \cdots dw_7 \, \delta(x+y+z+w_4+\cdots+w_7-1) \\
&   &
~~~~~~~~~~~~~~~~~~~~~~~~~~~~
\frac{1-y-z}{[(1-x-y-z) + (y+z)M^2\Lambda^{-2} - (1-y-z)(y+z)m^2\Lambda^{-2}]^4} \\
&   &
{} +
\mbox{lower order}.
\end{eqnarray*}
Using a standard integration identity, we get
\begin{eqnarray*}
F_2(0) 
& = &
\frac{\alpha}{2\pi} \cdot \frac{8Mm }{\Lambda^2} (f_1 + f_2 + f'')^2
\int_0^1 dx \int_0^{1-x} dy \int_0^{1-x-y} dz \\
&   &
~~~~~~~~~~~~~~~~~~~~~~~~~~~~
\frac{(1-y-z)(1-x-y-z)^3}
{[(1-x-y-z) + (y+z)M^2\Lambda^{-2} - (1-y-z)(y+z)m^2\Lambda^{-2}]^4} \\
&   &
{} +
\mbox{lower order} \\
& = &
\frac{\alpha}{2\pi} \cdot \frac{8Mm }{\Lambda^2} (f_1 + f_2 + f'')^2
I_1(m,M,\Lambda) +
\mbox{lower order},
\end{eqnarray*}
where
\[
I_1(m,M,\Lambda) =
\int_0^1 \int_0^{1-x} \int_0^{1-x-y} dx dy dz \,
\frac{
(1-y-z)(1-x-y-z)^3
}
{
[-(y+z)(1-y-z)m^2\Lambda^{-2} + (y+z)M^2\Lambda^{-2} + (1-x-y-z)]^4
}.
\]

We need to evaluate $F_2(0)$ in the regime $m \ll M \leq \Lambda$.
Since the integrand is a function of $y+z$, we can set $t = y+z$ and
simplify, using the fact that the Jacobian of the transformation
$(y,z)$ maps to $(y,t)$ is $1$, to get  
\begin{eqnarray*}
\lefteqn{
F_2(0)
} \\
& \approx &
\frac{\alpha}{2\pi} \cdot \frac{8Mm }{\Lambda^2} (f_1 + f_2 + f'')^2
\int_0^1 dx \int_0^{1-x} dt \int_0^t dy \,
\frac{(1-t)(1-x-t)^3}{[(1-x-t) + t M^2\Lambda^{-2}]^4} +
\mbox{lower order}.
\end{eqnarray*}

After some simplification, this expression becomes
\begin{eqnarray*}
\lefteqn{
F_2(0)
} \\
& = &
\frac{\alpha}{2\pi} \cdot \frac{8Mm }{\Lambda^2} (f_1 + f_2 + f'')^2
\int_0^1 dt \\
&         & 
~~~
\left(
\frac{\frac{t}{2} - \frac{t^2}{3}}{1 - t + t M^2 \Lambda^{-2}} +
\frac{3t^2(1-t) M^2 \Lambda^{-2}}{1 - t + t M^2 \Lambda^{-2}} -
\frac{3 t^3 (1-t) M^4\Lambda^{-4}}{2 (1 - t + t M^2 \Lambda^{-2})^2} +
\frac{t^4(1-t) M^6\Lambda^{-6}}{3 (1 - t + t M^2 \Lambda^{-2})^3} -
\frac{11 t(1-t)}{6}
\right) \\
&   &
{} +
\mbox{lower order},
\end{eqnarray*}
which has already been stated earlier.

\section*{Acknowledgements}
I am grateful to Urjit Yajnik and S. Umasankar for encouragement and
many useful discussions.

\end{document}